\documentclass[12pt]{iopart}
\pdfoutput=1
\expandafter\let\csname equation*\endcsname\relax
\expandafter\let\csname endequation*\endcsname\relax
\usepackage[numbers,sort&compress]{natbib}
\usepackage{amsmath,amssymb,eucal,graphicx,bm}
\usepackage{subfigure}
\usepackage{float}
\usepackage{xcolor}
\def\ltwid{\mathrel{\raise.3ex\hbox{$<$\kern-.75em\lower1ex\hbox{$\sim$}}}}
\def\gtwid{\mathrel{\raise.3ex\hbox{$>$\kern-.75em\lower1ex\hbox{$\sim$}}}}

\usepackage{appendix}
\usepackage[breakwords]{truncate}
\usepackage{lastpage}
\usepackage{hyperref}
\usepackage{amsfonts}
\usepackage[section]{placeins}
\usepackage{color}
\usepackage{url}

\usepackage{breakurl}
\usepackage{enumitem}
\usepackage{mathtools}

\begin{document}

\title{Fixation and Fluctuations in Two-Species Cooperation}

\author{Jordi Pi\~nero\footnote{jordi.pinero@upf.edu}, S. Redner\footnote{redner@santafe.edu} and Ricard Sol\'e\footnote{ricard.sole@upf.edu}}
\address{$^{1}$ICREA-Complex Systems  Lab, Universitat Pompeu Fabra, 08003 Barcelona, Spain}
\address{$^{2}$Institut de Biologia Evolutiva (CSIC-UPF), Psg Maritim Barceloneta, 37, 08003 Barcelona, Spain}
\address{$^{3}$Santa Fe Institute, 1399 Hyde Park Road, Santa Fe NM 87501, USA}

\begin{abstract}

  Cooperative interactions pervade in a broad range of many-body populations,
  such as ecological communities, social organizations, and economic webs.
  We investigate the dynamics of a population of two equivalent species A and
  B that are driven by cooperative and symmetric interactions between these
  species.  For an isolated population, we determine the probability to reach
  fixation, where only one species remains, as a function of the initial
  concentrations of the two species, as well as the time to reach fixation.
  The latter scales exponentially with the population size.  When members of
  each species migrate into the population at rate $\lambda$ and replace a
  randomly selected individual, surprisingly rich dynamics ensues.
  Ostensibly, the population reaches a steady state, but the steady-state
  population distribution undergoes a unimodal to trimodal transition as the
  migration rate decreases below a critical value $\lambda_c$.  In the
  low-migration regime, $\lambda<\lambda_c$, the steady state is not truly
  steady, but instead strongly fluctuates between near-fixation states, where
  the population consists of mostly A's or of mostly B's.  The characteristic
  time scale of these fluctuations diverges as $\lambda^{-1}$.  Thus in spite
  of the cooperative interaction, a typical snapshot of the population will
  contain almost all A's or almost all B's.
\end{abstract}


\maketitle

\section{Introduction}

Competitive interactions have played a prominent role in the literature of
ecological and evolutionary dynamics, as well as in economics and
sociology~\cite{goel1971volterra,murray2007mathematical,may2019stability}. Resource
limitations and their impact in defining the outcome of competition among
species has shaped a large part of evolutionary thinking.  A counterpoint to
competition is cooperativity in which there are positive interactions and
feedback loops between species.  These mechanisms have received increasing
attention recently~\cite{may2008ecology,bronstein2015mutualism}. In fact, it
is the presence of cooperative interactions, where positive reciprocal
exchanges are at work, that appear to drive innovations in evolution and also
maintain biodiversity in Nature~\cite{bronstein2015mutualism}.

\begin{figure*}[ht]
\begin{center}
\includegraphics[width= 12 cm]{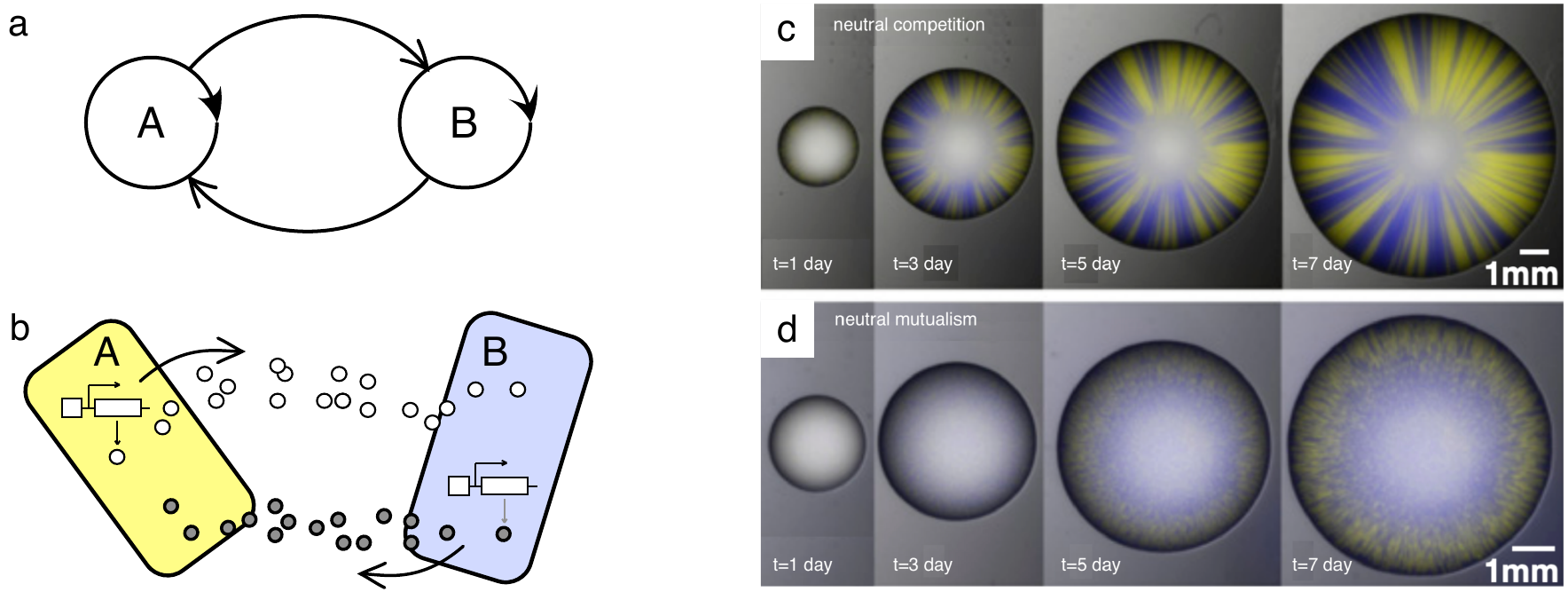}
\caption{In pairwise cooperation (a) the replication (closed arrows) of a
  given species A requires the help of B and vice versa.  This feedback
  occurs, for example, when two types of bacteria A and B each lack a
  metabolite for reproduction that is supplied by the other species (small
  circles in (b)).  Such cooperative feedbacks are commonplace and help
  maintaining diversity.  Experimental set-ups using engineered bacteria
  (c-d) reveal the marked difference between competitive and cooperative
  interactions.  In (c), two equal competitors grow on a Petri dish and
  locally exclude each other, as shown by the growing bands that indicate the
  presence of only one strain.  With cooperativity (d), the mutual dependency
  drives both strains to persist and mix.}
\label{fig:expt}
\end{center}
\end{figure*}

Cooperation, or mutualism, has been part of mathematical models of
populations since the formulation of Lotka-Volterra
equations~\cite{murray2007mathematical} and a variety of statistical physics
models of human cooperation~\cite{perc2016phase,perc2017statistical}.  In its
most abstract form, two species (for example, $A$ and $B$ in
Fig.~\ref{fig:expt}(a)) ``help'' each other by means of a mutual positive
interaction; in some cases, both partners completely rely on one another for
survival.  This feature underlies the two-species system in
Fig.~\ref{fig:expt}(b), where a given species requires the other to replicate
because each species needs a molecule that is produced by the partner
species.

Recent experimental studies have shown that such cooperative populations can
in fact be engineered. By following the scheme in Fig.~\ref{fig:expt}(b), it
is possible to create a completely symmetric pairwise dependence and make
these mixed populations grow on a Petri
dish~\cite{mitri2016resource,nadell2016spatial,muller2014genetic}.
Figure~\ref{fig:expt}(c) shows the outcome of symmetric competition when each
strain is marked with a different fluorescent protein: each strain locally
out competes the other, thereby generating stripes of segregated domains.
The cooperatively interacting population, on the other hand, constrains both
species to remain in proximity, leading to a well-mixed population
(Fig.~\ref{fig:expt}(d)).  These simple engineered, or synthetic, bacterial
populations, which can be tuned so that they become virtually symmetric,
allow one to explore the fundamental dynamical features of interacting living
consortia and also study the impact of
stochasticity~\cite{shou2007synthetic,amor2017spatial,rodriguez2019bottom}.

In this paper, we present an analytic approach to understand the role of
stochasticity in a simple two-species stochastic model of cooperation.  This
model represents a special case of evolutionary game
dynamics~\cite{nowak2006evolutionary,antal2006fixation,altrock2009fixation,black2012mixing},
with a specific and particularly simple form of the payoff matrix.  We
emphasize that we are not treating a general ecological model, but rather a
simplified system in which only cooperative interactions occur.  This
scenario appears to be more relevant for the microbiome~\cite{
  nadell2010emergence,rakoff2016evolution,foster2017evolution}. Current advances in 
  microbial ecology involve experimental setups
with a small number of interacting species~\cite{friedman2017ecological,vega2018simple}. 
In this model, we treat both
closed and open populations, in which there is either no migration or a
finite rate of migration into the population, respectively.  We define
microscopic rules that incorporate both cooperativity, in which each species
helps the other, as well as neutrality, in which neither species is preferred
over the other.  We first determine the steady state of the population in the
absence of stochastic fluctuations. For a finite population, we then
incorporate stochasticity and determine the time until fixation is reached
for the situation where no migration can occur.  When migration is allowed
(with compensatory removal), the population now reaches a steady state;
however, the character of this steady state dramatically changes as function
of the migration rate.  For large migration rate, both species are present in
roughly equal abundances.  However, for a sufficiently small migration rate,
the population strongly fluctuates between consisting of nearly all A or all
B.  Thus a \emph{typical} realization of the population has a completely
different character that the \emph{average} population.  This change in
behavior is mirrored by a bimodal to trimodal transition in the shape of the
steady-state probability distribution of species abundances.

In Sec.~\ref{sec:2s}, we outline basic features of our two-species
cooperation model in the absence of migration.  We solve the model in the
mean-field approximation and then include the role of stochasticity due to
the finiteness of the population.  For a finite population, we determine the
fixation probability as a function of the initial population composition and
the time until fixation, where only a single species remains.  In
Sec.~\ref{sec:mig}, we incorporate migratory inflow, with compensatory
removal, so that the population size remains fixed and reaches a steady
state.  We discuss basic features of this steady state, including the
intriguing feature that the species abundances can exhibit huge fluctuations,
even though time-averaged properties are constant.  We give some concluding
remarks in Sec.~\ref{sec:disc}

\section{Two-species cooperation}
\label{sec:2s}

\begin{figure*}[ht]
\begin{center}
\includegraphics[width= 8 cm]{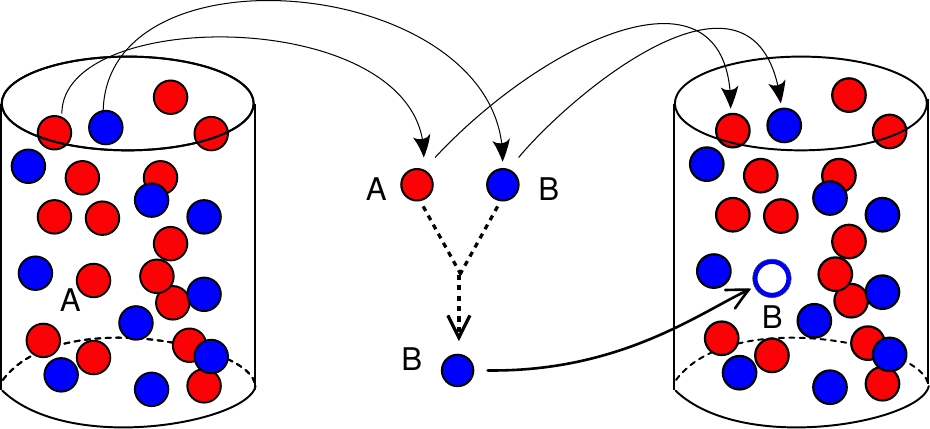}
\caption{The reaction step in two-species cooperation.  Two randomly selected
  particles happen to be from different species, namely A and B (red and
  blue).  One of them reproduces (B, blue), an event that is aided by the
  presence of the other (A, red).  The offspring replaces another randomly
  selected particle from the remainder of the population.  In the example
  shown, the newly generated B replaces an A.}
\label{fig:2}
\end{center}
\end{figure*}

We investigate a finite population of $N$ particles, with $n$ of species A
and $N-n$ of species B. The population undergoes repeated reaction events in
which each event consists of the following steps (Fig.~\ref{fig:2}):
\begin{enumerate}
  \itemsep -0.5ex
\item Pick a random pair of particles.  
\item If the pair is AB, one member of this pair reproduces; if the pair is
  AA or BB, nothing happens.
\item The newly reproduced offspring replaces one randomly selected particle
  in the remainder of the population.
\end{enumerate}

Thus interactions between members of different species are cooperative in
nature, while members of the same species are non interacting.  The
replacement step (iii) ensures that the total population remains fixed. The
lack of interactions between AA and BB pairs follows from the assumption of
strict mutualism, i.e., replication occurs if and only if both species are
present (as sketched in Fig.~\ref{fig:2}).  After each update, time is
incremented by $\frac{1}{N}$.  This time increment corresponds to each
particle undergoing, on average, steps (i)--(iii) in a single time unit.
While this dynamics manifestly conserves the total number of particles, the
composition of the population can change.  When the population consists
entirely of a single species---either all A's or all B's---there is no
further dynamics and the population fixates.

If an A reproduces in a single interaction, then with probability
$1-\frac{n}{N}\equiv 1-x$, the A offspring replaces a B and $n\to n+1$.
Conversely, with probability $x=\frac{n}{N}$, the A offspring replaces an
existing A so that $n$ does not change.   The probability $a_n$
at which $n\to n+1$ therefore is
\begin{subequations}
  \label{ab}
  \begin{align}
    \label{an}
  a_n = 2x(1-x)\times \tfrac{1}{2} \times (1-x) = x(1-x)^2\,.
\end{align}
Throughout, we use the variables $n$ and $x=\frac{n}{N}$ interchangeably.  In
\eqref{an}, the factor $2x(1-x)$ gives the probability a randomly selected
pair is AB, the factor $\frac{1}{2}$ gives the probability that the A in this
pair reproduces, and the factor $1-x$ gives the probability that the A
offspring replaces a B.  By the same reasoning, when a B reproduces in an
interaction, the probability at which $n\to n-1$ is
\begin{align}
  \label{bn}
  b_n = 2x(1-x)\times \tfrac{1}{2} \times x = x^2(1-x)\,.
\end{align}
Here the trailing factor $x$ in \eqref{bn} accounts for the probability that
the B offspring replaces an A.  Finally, the probability that the number of
A's and B's does not change is given by
\begin{align}
  x^2+(1-x)^2 +2x(1-x)\big[\tfrac{1}{2}x +\tfrac{1}{2}(1-x)\big]
  = 1-x(1-x)=1-a_n-b_n\,.
\end{align}
\end{subequations}
The terms $x^2+(1-x)^2$ give the probability to pick either an AA or BB
pair, for which no change in $n$ occurs.  For the last term, $2x(1-x)$ is
again the probability of picking an AB pair, while the factor in the square
brackets is the probability that the offspring (either A or B with
probability $\frac{1}{2}$) replaces its own kind so that $n$ does not change.

\section{Mean-Field Approaches}

\subsection{Rate  Equation}

Using the probabilities in Eqs.~\eqref{ab}, the rate equation for the average
number of A's is
\begin{subequations}
  \label{ndot}
\begin{equation}
  \label{ndotn}
   \dot n = N\left(a_n-b_n\right)=  Nx(1-x)(1-2x)\,,
\end{equation}
or equivalently,
\begin{equation}
  \label{xdot}
   \dot x =  x(1-x)(1-2x)\,.
\end{equation}
\end{subequations}
To keep the notation simple, $n$ and $x$ refer to average values in this
section; that is, we do not write angle brackets.  This rate equation has a
stable fixed point at $x=\frac{1}{2}$ and unstable fixed points at $x=0$ and
$x=1$.  The stability of the fixed point at $x=\frac{1}{2}$ arises because
the transition probabilities \eqref{ab} tend to reduce population imbalances.
Thus the steady state in this continuum description is a static population
that consists of equal densities of A's and B's.  That is, cooperativity
promotes diversity in the mean-field description.

The solution to the rate equation \eqref{xdot} may be straightforwardly
obtained by first performing a partial fraction decomposition:
\begin{align*}
  dt = \frac{dx}{x(1-x)(1-2x)} = dx\left(\frac{1}{x}-\frac{1}{1-x}+\frac{4}{1-2x}\right)\,,
\end{align*}
from which
\begin{align*}
  t &=\int_{x_0}^x  dy\left(\frac{1}{y}-\frac{1}{1-y}+\frac{4}{1-2y}\right)
                 = -4\ln \frac{x(1-x)(1-2x)}{x_0(1-x_0)(1-2x_0)}\,.
\end{align*}
We then obtain $x(t)$ by solving the resulting cubic equation.  For
$t\to\infty$, the limiting behavior is
\begin{align}
  x(t)\simeq \frac{1}{2}- 2\, x_0(1-x_0)(1-2x_0)\, e^{-t/4}\,,
\end{align}
so that the stable fixed point $x^*=\frac{1}{2}$ is approached exponentially
quickly in time.

\subsection{Master Equation and Its Moments}

In the stochastic dynamics where $n$ is a discrete variable, the true fixed
points are at $x=0$ and $x=1$.  Even though the fixed point at
$x^*=\frac{1}{2}$ is stable in the continuum limit, stochastic fluctuations
allow the population to explore the full state space and eventually get
trapped at either $x=0$ or $x=1$.  This behavior is analogous to the
extinction phenomena that arise, for example, in the logistic birth-death
process, $A\to 2A$ and $2A\to 0$, as well as other reactions of this
genre~\cite{elgart2004rare,kessler2007extinction,assaf2010extinction,assaf2017wkb}.
In these reactions, the rate equation predicts a steady population,
$N_{\rm s}$, which is determined by the balance between the birth and death
rates.  However, in the true stochastic dynamics, the population fluctuates
around $N_{\rm s}$, which actually is the \emph{quasi} steady-state value.
Ultimately, a sufficiently large fluctuation occurs that leads to extinction,
from which there can be no escape, with an extinction time that scales
exponentially in
$N_{\rm
  s}$~\cite{elgart2004rare,kessler2007extinction,assaf2010extinction,assaf2017wkb,krapivsky2010kinetic}.

To understand the stochastic dynamics for two-species cooperation, we study
$P_n(t)$, the probability that the population consists of $n$ A's and $(N-n)$ B's
at time $t$.  The time dependence of this probability distribution is given
by
\begin{align}
  \label{Pndot}
  \dot P_n &= N\Big[a_{n-1} P_{n-1} +b_{n+1} P_{n+1} -\big(a_n+b_n\big) P_{n}\Big]\,.
\end{align}
When the number of particles $N$ is small, the set of equations \eqref{Pndot}
can be straightforwardly solved.  For the initial condition of equal numbers
of A's and B's, both $P_0(t)$ and $P_N(t)$ approach $\frac{1}{2}$ as
$t\to\infty$, while all the other $P_n(t)$ vanish exponentially quickly in
time.  This direct approach quickly becomes tedious as $N$ increases,
however, and to gain insight into the long-time dynamics for general $N$, it
is useful to study low-order moments of $P_n$.  From Eq.~\eqref{Pndot} and
using $a_n$ and $b_n$ from Eqs.~\eqref{ab}, the first moment obeys
\begin{subequations}
\begin{align}
  \label{mom1}
  \langle \dot x\rangle = \frac{1}{N}\sum_n n\dot P_n
  &= \sum_{1\leq n\leq N}\left\{n\,a_{n-1} P_{n-1} +n\,b_{n+1} P_{n+1}
             -n\,\big(a_n+b_n)    P_{n}\right\}\nonumber\\
  &= \sum_{1\leq n\leq N}\left\{(n+1)\,a_n P_{n} +(n-1)\,b_n P_{n}
    -n\,\big(a_n+b_n\big)  P_{n}\right\}\nonumber\\
   &= \sum_{1\leq n\leq N}(a_n-b_n) P_{n}
  =\big\langle x(1-x)(1-2x)\big\rangle\,.
\end{align}
Here we now explicitly write angle brackets to denote average values.  Under
the assumption of no correlations, that is,
$\langle x^k\rangle=\langle x\rangle ^k$, \eqref{mom1} reproduces the rate
equation \eqref{xdot}.

Similarly, the equation of motion for the second moment is
\begin{align}
  \label{mom2}
  \langle \dot x^2\rangle = \frac{1}{N^2}\sum_n n^2\dot P_n
  &= \frac{1}{N}\sum_{1\leq n\leq N}\left\{n^2a_{n-1} P_{n-1} +n^2\,b_{n+1} P_{n+1}
       -n^2\,\big(a_n+b_n\big)  P_{n}\right\}\nonumber\\
  &= \frac{1}{N}\sum_{1\leq n\leq N}\left\{(n+1)^2\,a_n P_{n} +(n-1)^2\,b_n P_{n}
     -n^2\,\big(a_n+b_n\big) P_{n}\right\}\nonumber\\
             &=\frac{1}{N}\langle x(1-x)\rangle+ 2\langle x^2(1-x) (1-2x)\rangle\,.
\end{align}
\end{subequations}
It is more convenient to express \eqref{mom1} and \eqref{mom2} in terms of
$z\equiv 2x-1$, which lies in the range $[-1,1]$.  Doing so, we obtain
\begin{align}
  \label{2-moms}
\begin{split}
\langle \dot z\rangle &=-\tfrac{1}{2}\langle z(1-z^2)\rangle\\
\langle \dot z^2\rangle&=\big\langle (1-z^2)\big(\tfrac{1}{N}-z^2)\big\rangle\,,
\end{split}
\end{align}
which are both symmetric about $z=0$.  If we make the assumption of no
correlations, that is, $\langle z^k\rangle=\langle z\rangle^k$, then the
first equation reproduces the result that $z=0$ is a stable fixed point.  The
second equation then predicts that the width of the distribution initially
grows and eventually ``sticks'' at the value $\sqrt{N}$.  To check this
point, we numerically integrated Eqs.~\eqref{Pndot} for small values of $N$.
From the resulting solution, we find that the width of the probability
distribution initially grows with time and later approaches a nearly fixed
value that is proportional to $\sqrt{N}$.  However, at very long times, there
is slow leakage of the probability distribution to the true stochastic fixed
points at $z=\pm 1$.  Thus the probability distribution eventually approaches
two delta-function peaks at these fixed points.  This behavior cannot be
captured by low-order moment equations, such as \eqref{2-moms}.  Instead, we
need to study the full stochastic dynamics; this is done in the following
section.

\section{Fixation Probability and Fixation Time}

We now turn to two quantities of primary interest in the stochastic dynamics,
namely, (i) the fixation (or exit) probability $E_n$, and (ii) the fixation
time $T_n$.  The fixation probability $E_n$ is defined as the probability
that a population of size $N$ that initially contains $n$ particles of type A
reaches the static fixation state of all A's.  We use the backward Kolmogorov
equation~\cite{van1992stochastic,redner2001guide} to compute the fixation
probability.  In this approach, $E_n$ satisfies the recursion
\begin{align}
\label{En}
  E_n= a_n E_{n+1} + b_n E_{n-1}  +\big(1-a_n-b_n\big)E_n\,.
\end{align}
Since the process renews itself after each event, we can express the fixation
probability from the state that contains $n$ A's in terms of the
appropriately weighted average of the fixation probabilities after a single
step to the states $n-1$, $n$, and $n+1$.  The weights are merely the hopping
probabilities to these respective states.  Equation~\eqref{En} is subject to
the boundary conditions $E_0=0$ and $E_N=1$.  The first condition corresponds
to the impossibility of reaching a population of all A's if the initial state
contains no A's, while the second condition corresponds to the initial state
coinciding with the desired final state of all A's

\begin{figure}[ht]
\begin{center}      
  \subfigure[]{\includegraphics[width=0.425\textwidth]{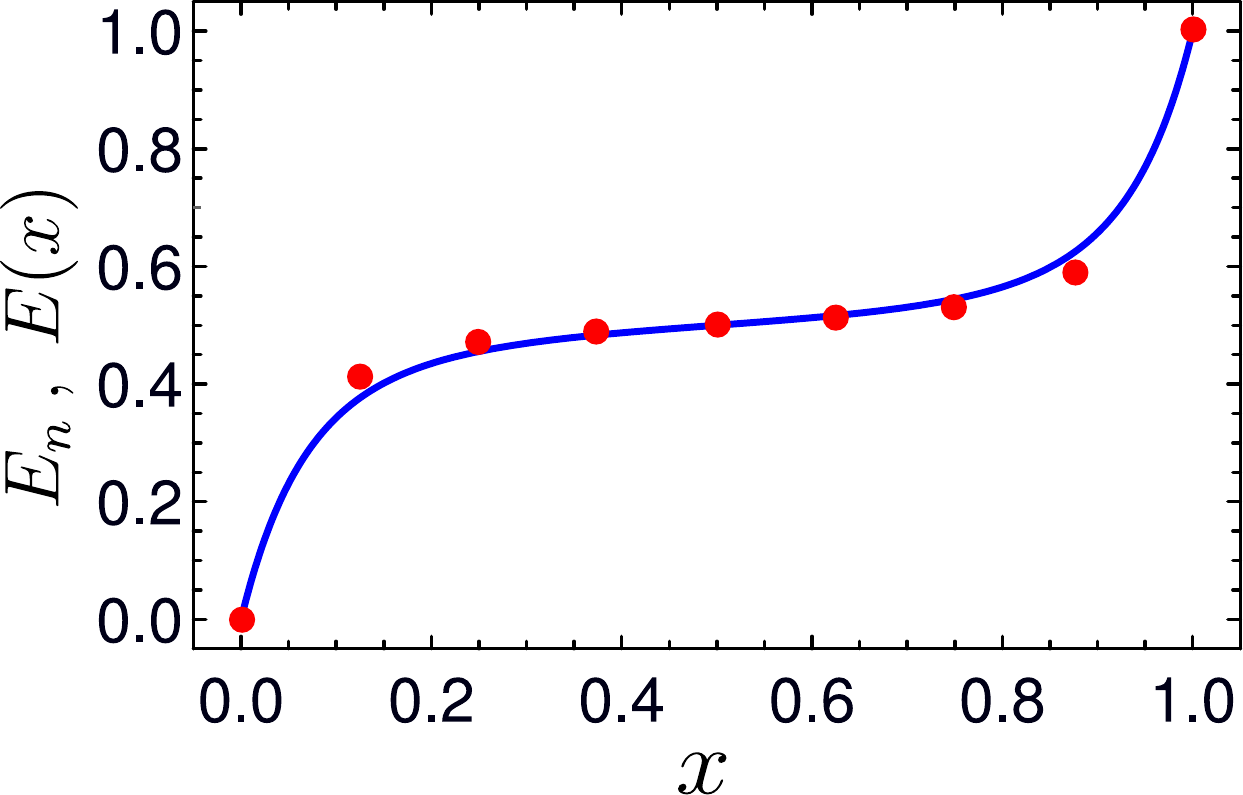}}\qquad
   \subfigure[]{\includegraphics[width=0.425\textwidth]{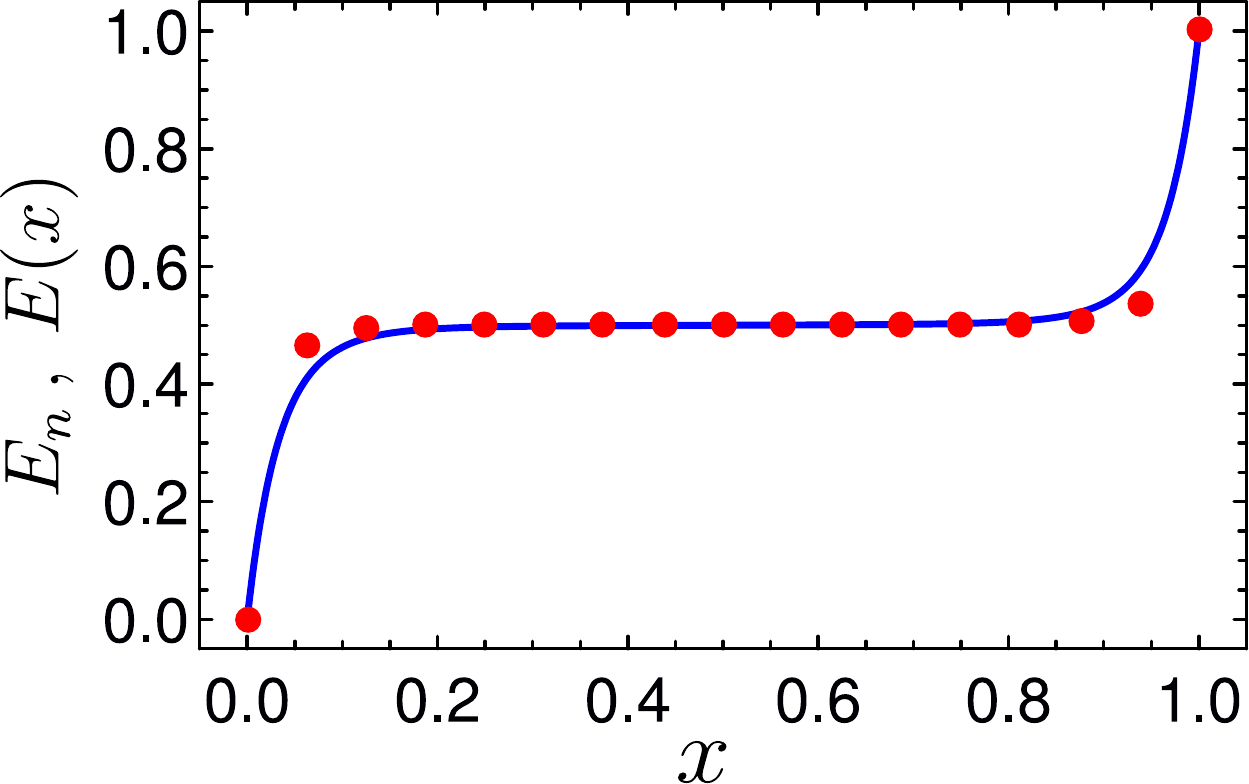}}
   \caption{Dependence of the discrete and continuum fixation probabilities,
     $E_n$ and $E(x)$, versus $x$ for the cases $N=8$ and 16.  The smooth
     curves represent $E(x)$ from Eq.~\eqref{Ex-soln} and the dots represent
     simulation results. }
\label{En-Ex}
\end{center}
\end{figure}

The solution to \eqref{En} is (see \ref{app:En} for the calculational
details)
\begin{align}
\label{En-soln}
  E_n = \sum_{m=0}^{n-1} \left[\binom{N-1}{m}\right]^{-1} \,\Big/
  \sum_{m=0}^{N-1} \left[\binom{N-1}{m}\right]^{-1} \,.
\end{align}
Neither of these sums has a closed form, but for $N\to\infty$ the denominator
approaches 2~\cite{stack}.  For $N\gg 1$, $E_n$ and its continuum counterpart
$E(x)$ (see also \ref{app:En}) are nearly independent of $x=\frac{n}{N}$ when
$x$ is not close to 0 or 1.  Figure~\ref{En-Ex} shows this dependence of
$E_n$ on $n$.  Also shown are the corresponding results from discrete
simulations of the fixation process.  Simulations are carried out by setting a finite size ($N$) array
of particles with two possible states. Each iteration the 
particles interact following rules (ii) and (iii) with the interaction rates given by
expressions~\eqref{an} and~\eqref{bn}. Time is updated by $\Delta t= \left[N \left(a_n+b_n\right)\right]^{-1}$ 
after each iteration. 

The anti-sigmoidal shape of $E_n$ arises from the underlying drift that tends
to drive any initial population towards $x=\frac{1}{2}$.  Eventually, a large
and rare stochastic fluctuation causes the population to escape this
effective potential well and reach fixation.  This anti-sigmoidal shape also
strongly contrasts with the Moran process~\cite{moran1962statistical}, which
is symmetric (neutral), but non-cooperative.  Here an AB pair equiprobably
converts to either AA or to BB.  As a result of this lack of cooperativity,
the fixation probability in the strictly neutral Moran process is simply the
linear function
$E(x)=x$~\cite{moran1962statistical,kimura2020neutral,ewens2012mathematical,redner2001guide}.

\begin{figure}[ht]
\begin{center}      
 \subfigure[]{\includegraphics[width=0.425\textwidth]{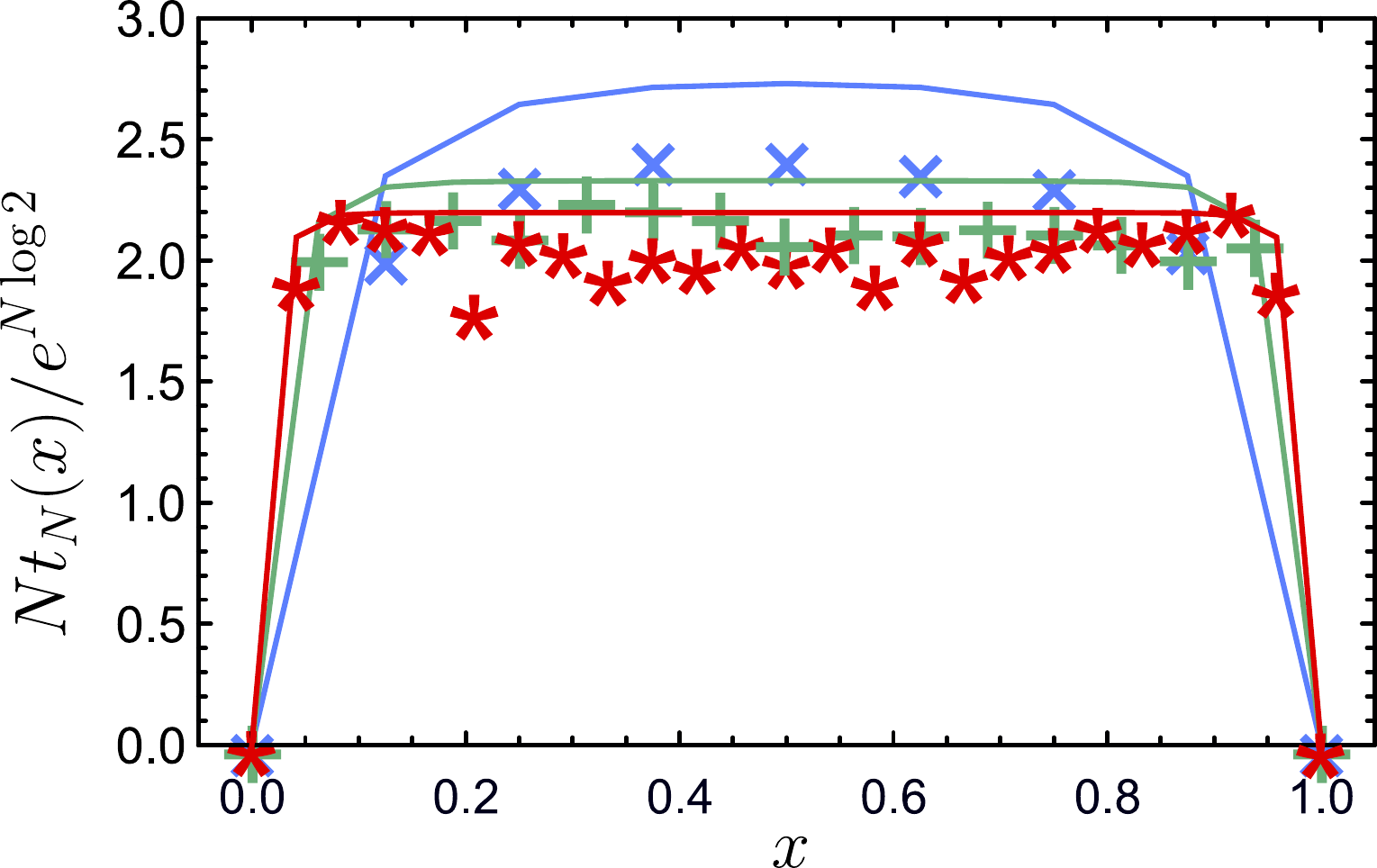}}\qquad
 \subfigure[]{\includegraphics[width=0.425\textwidth]{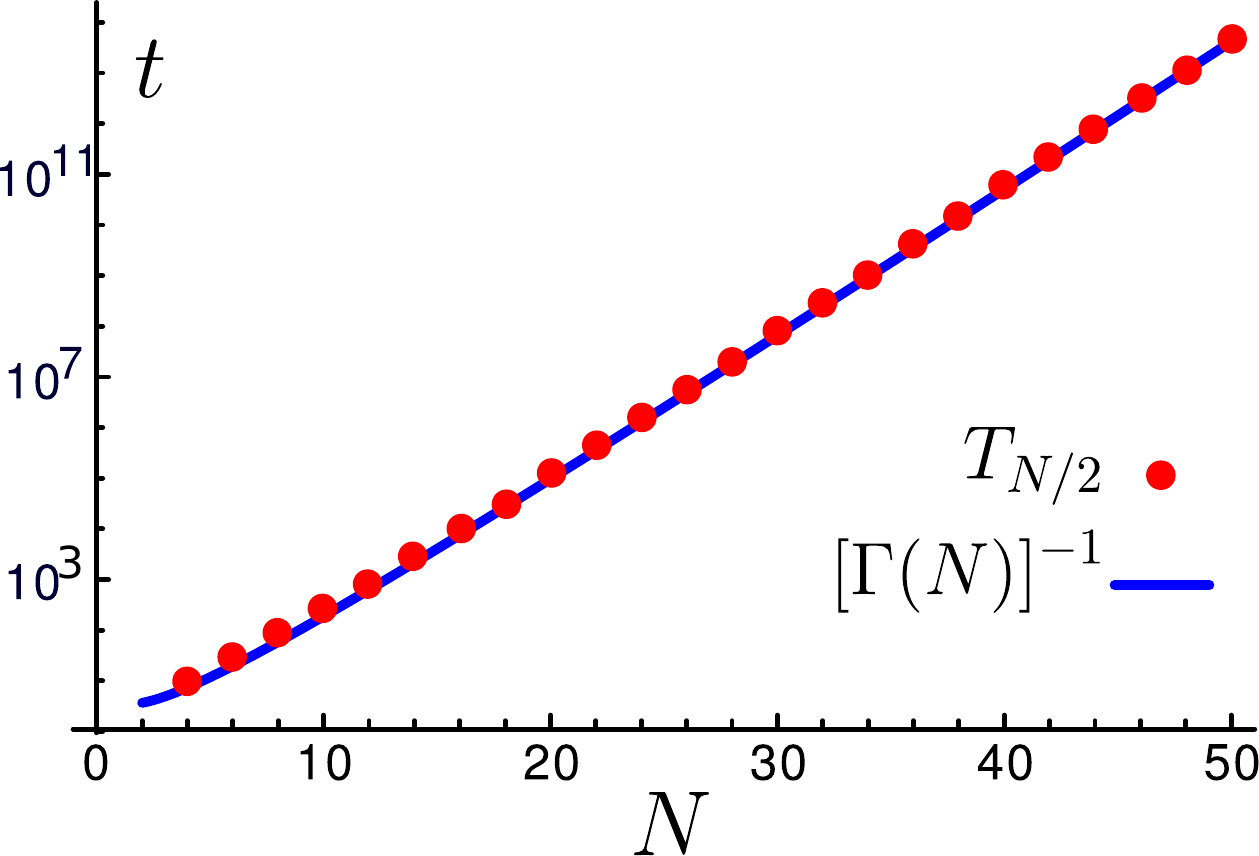}}
  
 \caption{(a) Dependence of the fixation time $T_n$ versus $x=\frac{n}{N}$ using a data-collapse scheme by resetting the scale $T_n \to Ne^{-N\log2} T_n$ for each $N$ value. This scaling factor corresponds to the prediction made in Eq.~\eqref{GammaN}. The solid lines follow the predictions obtained from  Eq.~\eqref{tn-final}. The plot marks \{$\times$, + and $*$\} correspond to simulated average fixation times for $N=8 , \ 16$ and $24$ in blue, green and red colors, respectively. Each data point is obtained by averaging $10^3$ simulated fixation processes at corresponding values of $n$ and $N$. (b) Dependence of the fixation time from the symmetric initial state, $T_{N/2}$ (red dots) computed by~\eqref{tn-final} and the WKB prediction for the inverse leakage rate from the quasi-steady state (blue line), following~\eqref{GammaN}. }
\label{fig:tn}
\end{center}
\end{figure}

We now investigate the fixation time $T_n$, which is defined as the average
time for the population of $N$ particles to first reach \emph{either} of the
two fixation states, $n=0$ or $n=N$, when the population initially contains
$n$ A's.  Within the same backward Kolmogorov framework as that used for the
fixation probability, the fixation time
satisfies~\cite{van1992stochastic,redner2001guide}
\begin{align}
  \label{Tn}
  T_n&= a_nT_{n+1}+b_nT_{n-1}+\big(1-a_n-b_n\big)T_{n}+\delta t\,.
\end{align}
Again $T_n$ may be expressed as the appropriately weighted average of the
fixation time after a single hopping event to the states $n-1$, $n$, and
$n+1$, plus the time $\delta t=\frac{1}{N}$ required for this single step.
The latter corresponds to each particle being updated once, on average, in a
single time unit.  The equation for $T_n$ is subject to the boundary
conditions $T_0=T_N=0$; namely, if the population starts in a fixation state,
the time to reach this state is 0.  The result for the fixation time is (see
\ref{app:t})
\begin{align}
  \label{tn-final}
  T_n   = E_n   \sum_{m=1}^{N-1}  Q_m- \sum_{m=1}^{n-1} Q_m\,,
\end{align}
where
\begin{align*}
  Q_{n} \equiv  \alpha_{n}+r_{n}\,\alpha_{n-1}+r_{n}\,r_{n-1}\,\alpha_{n-2}+\cdots+
  r_{n}\, r_{n-1}\cdots r_2\, \alpha_1\,,
\end{align*}
with $Q_0=0$, $r_n=b_n/a_n$, and $\alpha_n\equiv \delta t/a_n$.

It does not seem possible to reduce \eqref{tn-final} to a compact form, but
the main feature of this exact expression is that the fixation time scales
exponentially in $N$ and is nearly independent of $n$ (or, equivalently,
$x$), except for $n$ close to 0 or to $N$ (Fig.~\ref{fig:tn}).  The
exponential dependence on $N$ again arises because of the existence of an
effective potential well, whose depth grows linearly with $N$, which draws
the population toward the state $x=\frac{1}{2}$.  The near independence of
the fixation time on the initial condition is a consequence of the population
being drawn toward the bottom of this potential well, where the
concentrations of A and B are equal.  As a result, the value of the fixation
time for any initial value of $x$ is close to the fixation time when the
population starts from the bottom of the potential well at $x=\frac{1}{2}$.

It is possible, however, to obtain an analytical expression for the average
fixation time by the WKB
method~\cite{elgart2004rare,kessler2007extinction,assaf2010extinction,assaf2017wkb}.
The idea of this approach is that the probability distribution settles into a
quasi-steady state that assumes an exponential large-deviation form.  From
Eq.~\eqref{Pndot}, there is a slow leakage from this quasi-steady state to
the fixation state whose rate, $\Gamma(N)$, is given by
\begin{align}
  \label{fixation_rate}
  \Gamma(N)\,\delta t=b_{1}\widetilde{P}_1+a_{N-1}\widetilde{P}_{N-1} = 2b_1\widetilde{P}_1 \ ,
\end{align}
i.e., the flux from states that are one step away from fixation to the
fixation states.  Here the tilde denotes the steady-state distribution and we
also use the symmetry $n\leftrightarrow N-n$.  We then identify the inverse
of this leakage rate with the fixation time.

We obtain an approximate equation for the continuum probability distribution
$\widetilde{P}_n\to \widetilde{P}(x)$ by setting the time derivative in the
master equation \eqref{Pndot} to zero, and writing $n\pm 1$ as $x\pm\delta x$
to give
\begin{align}
\label{quasisteady_process}
  a(x-\delta x)\widetilde{P}(x\!-\!\delta x) + b(x+\delta x) \widetilde{P}(x\!+\!\delta x)
  =\left[a(x)+b(x)\right]\widetilde{P}(x) \ .
\end{align} 
We now assume that $\widetilde{P}(x)$ has the exponential form
$\widetilde{P}(x)\sim e^{S(x)/\delta x}=e^{NS_0(x)+S_1(x)+\cdots}$ and
substitute this form into \eqref{quasisteady_process} to give (up to $O(1)$)
\begin{align}
\label{S0_S1}  
  S_0(x)=\int^x dz \log\left[\frac{a(z)}{b(z)}\right] \ , \quad S_1(x)=-\frac{1}{2}\log\left[a(x)b(x)\right]\,.
\end{align}

Now using \eqref{an} and \eqref{bn} for $a(x)$ and $b(x)$, we have
\begin{align}
\tilde{P}(x)\sim\frac{e^{NS_0(x)}}{x^{3/2}(1-x)^{3/2}}\ ,
\end{align}
with $S_0(x)=-x\log x -(1-x)\log (1-x)$. Note that the action $S_0(x)$ is peaked at the quasi-state state
$x=\frac{1}{2}$.  We normalize $\widetilde{P}(x)$ by using the Laplace method
for $N\to\infty$~\cite{bender2013advanced},
\begin{align*}
  \int^1_0 \frac{e^{NS_0(x)}}{x^{3/2}(1-x)^{3/2}}\,dx\approx \sqrt{\frac{32}{N}}\; e^{N\log 2}\ ,
\end{align*}
so that
\begin{align}
\label{WKB_quasisteady_solution}
\widetilde{P}(x)\simeq\frac{N}{\sqrt{32\pi}}\;
\frac{e^{N\left[-x\log x -(1-x)\log (1-x)-\log 2 \right]}}{x^{3/2}(1-x)^{3/2}} \ .
\end{align}

We now compute the fixation rate $\Gamma$ by substituting
\begin{align*}
  \widetilde{P}\big(\tfrac{1}{N}\big)\simeq\frac{N^3}{\sqrt{32 \pi}}\;e^{1-N\log2}  \quad \text{and} \quad b\big(\tfrac{1}{N}\big)\simeq \frac{1}{N^2}\,,
\end{align*}
into Eq.~\eqref{fixation_rate} to give
\begin{align}
  \label{GammaN}
\Gamma(N)=\frac{Ne}{\sqrt{ 8 \pi}}\; e^{-N\log 2} \ .
\end{align}
We now identify the inverse of this rate with the average fixation time.  As
shown in Fig.~\ref{fig:tn}(b), this inverse rate accurately matches the
simulation data for the fixation time.

\section{Two-species cooperation with migration}
\label{sec:mig}

We now incorporate migration into the dynamics, in which particles of either
species migrate into the population at the same fixed rate $\lambda$, and
each new particle replaces a randomly selected existing particle.  Because
migration is accompanied by replacement, the population size remains fixed,
which is the physically most relevant case.  Now the population is driven to
a steady state rather than to fixation and we want to understand the nature
of this steady state.

\subsection{Probability Distribution}

For a population that consists of $n$ A's and $(N-n)$ B's, suppose that the
migrant is an A.  With probability $\frac{1}{2}(1-x)$, the A migrant replaces
a B and $n\to n+1$, while with probability $\frac{1}{2}x$, the A migrant
replaces an A, and the composition of the population remains the same.
Similar reasoning applies when the migrant is a B.  As a result of a
migration event, the average change in the number of A's is
$\frac{1}{2}(1-x)- \frac{1}{2}x$.  The rate equation for $n$ now is (compare with
Eq.~\eqref{ndot})
\begin{equation}
  \label{ndoti}
\langle \dot n\rangle=  N(1-\lambda)\left[ x(1-x)(1-2x)\right] +\tfrac{1}{2}N\lambda(1-2x)\,.
\end{equation}
For $\lambda>0$, $x=0$ and $x=1$ are no longer fixed points and only the
remaining fixed point at $x=\frac{1}{2}$ is stable.  In the absence of
fluctuations, the population is thus driven to a steady-state distribution,
$P_n(t\to\infty)$, that is peaked about $x=\frac{1}{2}$.  Because there is no
absorbing state in the stochastic dynamics, we might anticipate a similar
behavior for $P_n(t\to\infty)$ when stochasticity is accounted for.  We will
show, however, that within the Fokker-Planck approximation the steady-state
distribution can either be unimodal or trimodal in shape and the latter case
corresponds to a steady state that is not truly steady.


The probability distribution $P_n$ is now governed by the master equation
  \begin{align}
    \label{ME-two}
  \dot P_n&  =N(1-\lambda)\left[ a_{n-1} P_{n-1} +b_{n+1} P_{n+1}
              -\big(a_n+b_n\big) P_{n}\right]\nonumber\\
           &\qquad\qquad +N\lambda\left[c_{n-1}P_{n-1}+ d_{n+1}P_{n+1}-
              \big(c_n+d_n\big)P_n\right]\,,
\end{align}
with hopping probabilities due to migration that are given by
\begin{align}
  \label{cd}
  c_n= \frac{1}{2}\left(1-\frac{n}{N}\right)\qquad\qquad
  d_n=\frac{1}{2}\frac{n}{N}\,.
\end{align}
We now determine the continuum probability distribution in the Fokker-Planck
approximation.  As we shall see, this continuum expression for the
probability distribution matches simulation data quite well, thus justifying
the Fokker-Planck approximation \emph{ex post facto} as a way to probe
steady-state properties.

In terms of $x=\frac{n}{N}$, $dx=\frac{1}{N}$, $P_n\to P(x)$, we expand
\eqref{ME-two} in a Taylor series up to second order.  This gives the
Fokker-Planck equation~\cite{gardiner1985handbook,van1992stochastic}
\begin{align}
  \label{Pt}
  P_t &
        =  -\Big\{\!(1\!-\!2x)\left[(1\!-\!\lambda)\,x(1\!-\!x)+\tfrac{\lambda}{2}\right]P(x,t)\!\Big\}_x
        +\frac{1}{2N}\Big\{\!\left[(1\!-\!\lambda)\,x(1\!-\!x)+\tfrac{\lambda}{2}\right]P(x,t)\!\Big\}_{xx}
        \nonumber\\[2mm]
      &\equiv - \big\{v(x) \, P(x,t)\big\}_x+ \big\{D(x)\,P(x,t)\big\}_{xx} \,,
\end{align}
where the subscripts denote partial derivatives.

The steady state is defined by solving this equation with the left-hand side
set to zero.  Integrating once gives $(D P)_x -v P = B$, where $B$ is a
constant.  We determine the constant by evaluating this equation at the
symmetry point $x=\frac{1}{2}$.  Because the probability distribution is
symmetric about $x=\frac{1}{2}$, $P_x(x\!=\!\frac{1}{2})=0$.  Moreover, at
$x=\frac{1}{2}$, $v =0$ and $D_x=0$, which implies that $B=0$.  Thus we only
need to solve $(D P)_x -v P = 0$, whose solution is
\begin{align}
  \label{Px}
P(x) &= C \,\exp\bigg\{\int^{x}dy\,\,
  \frac{v(y)-D_{y}(y)}{D(y)}\bigg\}=C \,\exp\bigg\{-\log D(x) + \int^{x}dy\,\,
  \frac{v(y)}{D(y)}\bigg\}\nonumber\\
  &= \frac{C}{D(x)}\; \exp\bigg\{\int^{x}dy\,\,2N(1-2y)\bigg\}
    \nonumber\\
 &=  C^{\prime} \, \left[\frac{1}{(1\!-\!\lambda)x(1\!-\!x)+\frac{\lambda}{2}}\right] \,\, e^{2Nx(1-x)}\,,
\end{align}
where the constant $C'$ is determined by normalization.

\begin{figure}[ht]
       \begin{center}
          \includegraphics[width=7 cm]{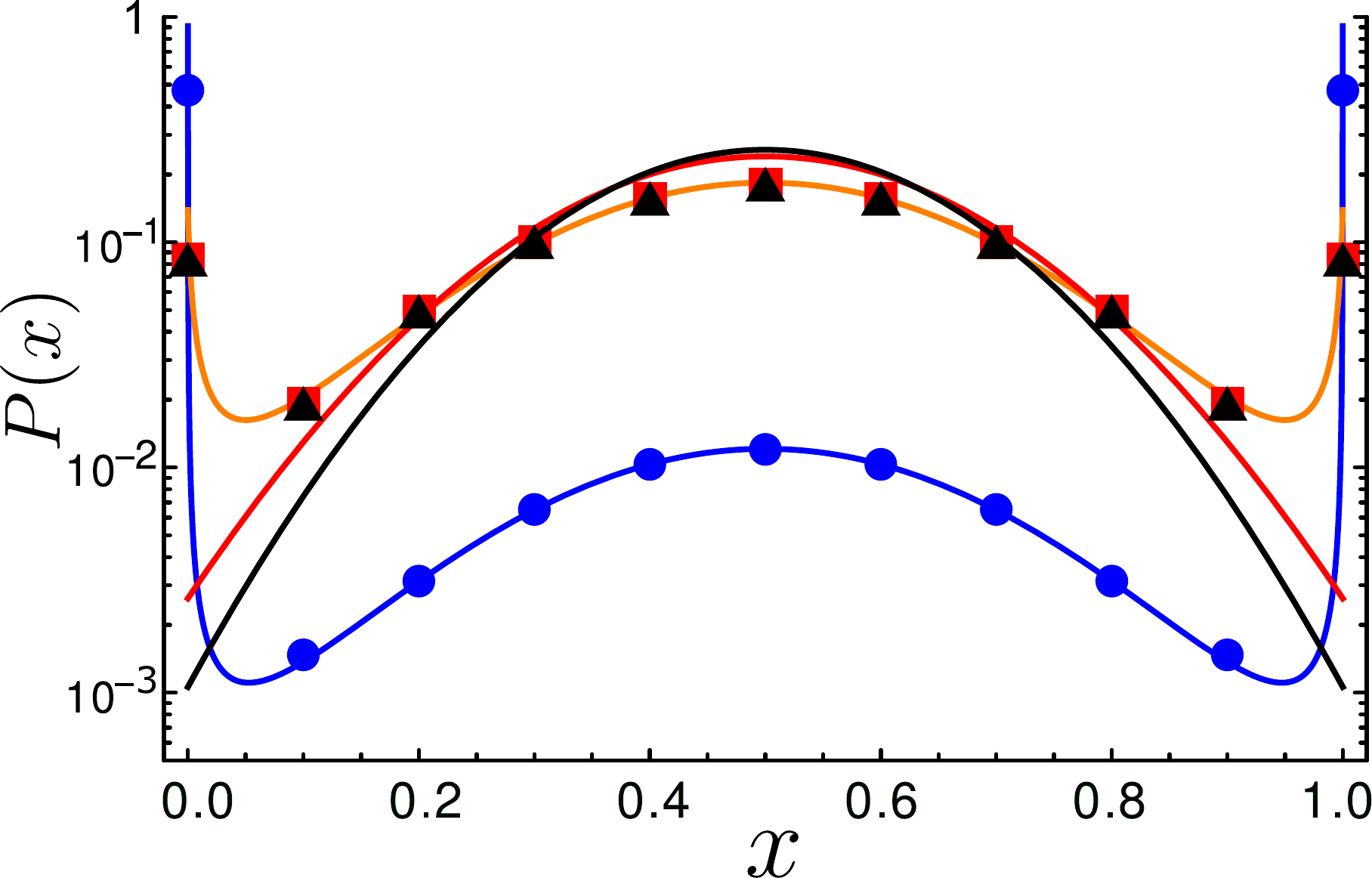}
          \caption{Steady-state probability distributions for $N=10$ on a
            semi-logarithmic scale for values of $\lambda/\lambda_c=10^{-4}, 10^{-2}, 10^0 \  \text{and} \ 10^1$ in blue, orange, red and black, respectively.  The  respective data
            points correspond to simulation results.}
  \label{Pss}
         \end{center}
\end{figure}

For $\lambda\to 0$, $P(x)$ is concentrated near $x=0$ and near $x=1$; these
peaks correspond to the near-fixation states.  Because of rare fluctuations,
however, the population stochastically switches between states where almost
all particles are of type A to states where almost all particles are of type
B.  Naively, one therefore anticipates that the steady-state distribution
should be bimodal, with a peak at each of the two near-fixation states.
Unexpectedly, there always remains a peak at $x=\frac{1}{2}$ (which may be
vanishingly small), so that the this distribution is \emph{trimodal} in the
small-$\lambda$ regime.  As $\lambda$ increases beyond a critical value, the
steady-state distribution undergoes a trimodal to unimodal transition
(Fig.~\ref{Pss}).

For fixed $N$, we determine the transition between trimodality and
unimodality by finding the point(s) where $P'(x)=0$.  This calculation gives,
after straightforward algebra,
\begin{align*}
    P'(x) \propto  (1-2x)\,\, e^{2Nx(1-x)}\,\,   \left[2N-\frac{(\lambda-1)}{D(x)^2}\right]\,,
\end{align*}
where again $D(x)$ is the diffusion coefficient defined by Eq.~\eqref{Pt}.
The leading factor of $1-2x$ equals 0 at $x=\frac{1}{2}$ and corresponds to
the extremum in the distribution at $x=\frac{1}{2}$.

However, there are additional extrema at the points where the factor in the
square brackets equals 0.  To determine these extrema, we first determine the
zero of this factor at $x=0,1$.  Thus we have the condition
$2N =(1-\lambda)/(\lambda/2)^2$.  Since we will find that $\lambda\ll 1$, we
also neglect $\lambda$ compared to 1 to give
\begin{align}
  \lambda_c= \sqrt{\frac{2}{N}}\,.
\end{align}
For $\lambda<\lambda_c$, the distribution $P(x)$ has three extrema.  To find
the location of the two secondary extrema for $\lambda \lesssim \lambda_c$,
it is convenient to now use the variable $y=x-\frac{1}{2}$ with
$y\to\pm \frac{1}{2}$, which corresponds to $x$ close to zero or to 1.  Now
the condition that the factor in the square brackets equals zero gives
\begin{align*}
2N\simeq  \frac{1}{(2\epsilon +\lambda/2)^2}\,,
\end{align*}
from which we obtain
\begin{align}
  \epsilon \simeq \tfrac{1}{4}(\lambda_c-\lambda)\qquad \text{for}\quad\lambda<\lambda_c
\end{align}

In the regime $\lambda<\lambda_c$, the distribution $P(x)$ is necessarily
trimodal because the point $x=\frac{1}{2}$ is always a local maximum of
$P(x)$.  To verify this point, we compute the second derivative of $P(x)$ at
$x=\frac{1}{2}$,
\begin{align*}
P''(x)= e^{2Nx(1-x)}\left[(1-2x)^2 - 2 + \frac{2(1-\lambda)}{D(x)^3} \;D'(x)\right]\,,
\end{align*}
which is indeed negative at $x=\frac{1}{2}$.

\subsection{Macroscopic Fluctuations in the Steady State}

\begin{figure}[ht]
\centerline{
  \includegraphics[width=9cm]{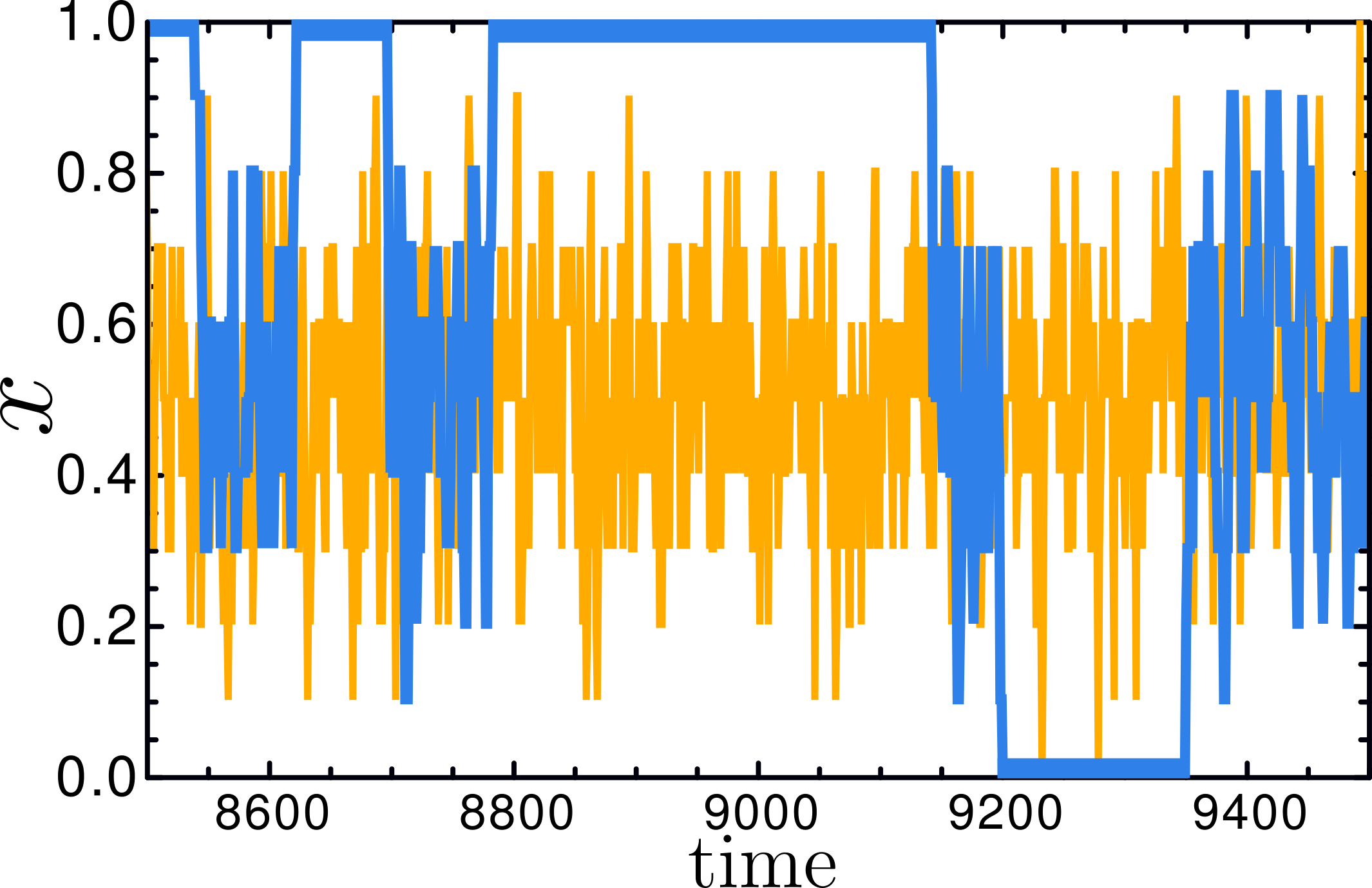}}
\caption{ Typical population trajectories in composition space for
  $\lambda/\lambda_c=2\times10^{-3}$ (blue) and $\lambda/\lambda_c=2$ (orange) for the
  case $N=10$.}
\label{fig:traj}
\end{figure}

An intriguing feature of 2-species cooperation with migration is that the
steady state is not strictly steady, especially when $\lambda$ is small
(Fig.~\ref{fig:traj}).  For $\lambda\ll\lambda_c$, substantial time ranges
exist during which little or no immigration occurs.  During these periods,
the population will tend to approach one of the fixation states.  Even if the
population does reach a state of all A's or all B's, immigration eventually
drives the population towards an equal number of A's and B's.  The competing
effects of cooperation and immigration therefore cause the population to
wander stochastically from one near-fixation state to the other, with returns
to the equal-concentration point $x=\frac{1}{2}$ controlled by the
immigration rate (Fig.~\ref{fig:traj}).  Related phenomenology occurs in
noisy voter
models~~\cite{fichthorn1989noise,considine1989comment,carro2016noisy,herrerias2019consensus}.
The extended time periods during which the population is close to fixation
corresponds to the large weight in the secondary peaks of the probability
distribution in Fig.~\ref{Pss}.  In contrast, when the immigration rate is
much larger than $\lambda_c$, the rapid inflow of equal numbers of A's and
B's ensures that the population contains roughly equal numbers of both
species.

One way to quantify these fluctuations is by the \emph{return time} $RT$,
which we define as the time interval between successive points where
$x=\frac{1}{2}$.  As suggested by Fig.~\ref{fig:traj}, this return time will
be short for $\lambda\gg\lambda_c$ and become long for
$\lambda\ll \lambda_c$.  The latter behavior is a harbinger of large
composition fluctuations in the population.  We can again use the backward
Kolmogorov approach to determine the $\lambda$ dependence of the return time.
Let $\tau_n$ denote the average time to reach the balanced state of equal
numbers of A's and B's when starting from a state where the number of A's
equals $n>\frac{1}{2}N$.  Then
\begin{align} RT=\frac{1}{N}+\tau_{1+N/2}\,.
\end{align}
That is, starting from the balanced state, the average return time to this
state is the average time for a single event, $\frac{1}{N}$, where the
population now consists of $\frac{N}{2}+1$ A's and $\frac{N}{2}-1$ B's, plus
the average time to reach the balanced state when starting from this
minimally imbalanced state.

\begin{figure}[ht]
\begin{center}
\includegraphics[width=0.5\textwidth]{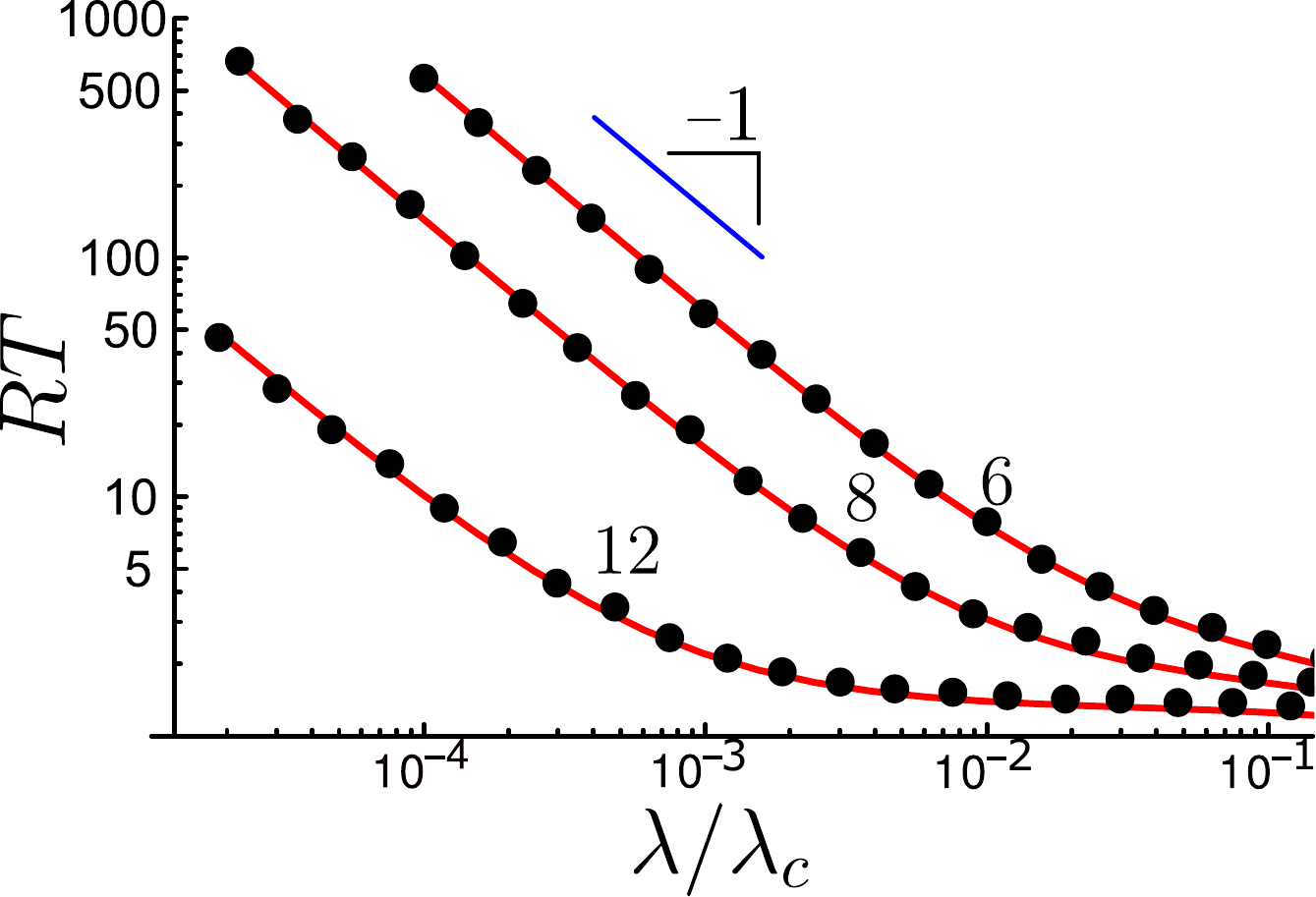}
\caption{Dependence of the return time $RT$ on $\lambda/\lambda_c$ for
  various values of $N$.}
\label{RT}
\end{center}
\end{figure}

In analogy with Eq.~\eqref{Tn} for the fixation time, the time $\tau_n$
satisfies the recursion
\begin{subequations}
\begin{align}
  \label{taun}
  \tau_n&= (1-\lambda)\left[a_n\tau_{n+1}+b_n\tau_{n-1}+\big(1-a_n-b_n\big)\tau_{n}\right]\nonumber\\
        &\qquad\qquad  +\lambda\left[c_n\tau_{n+1}+d_n\tau_{n-1}+\big(1-c_n-d_n\big)\tau_{n}\right]
          +\delta t\,,
\end{align}
with $a_n$, $b_n$ given by \eqref{an} and \eqref{bn}, and $c_n$, $d_n$ given
by \eqref{cd}.  We can rewrite this recursion in the canonical form of
Eq.~\eqref{Tn}:
\begin{align}
  \label{taup}
\tau_n= a_n'\tau_{n+1} +b_n'\tau_{n-1} +(1-a_n'-b_n')\tau_n+\delta t\,,
\end{align}
\end{subequations}
where $ a_n' =(1-\lambda)a_n+\lambda c_n$ and
$b_n'= (1-\lambda )b_n +\lambda d_n$.  This recursion is valid for
$\frac{1}{2}N<n<N$, while for $n=N$ the first term in \eqref{taup} is absent.
This missing term acts as the effective reflecting boundary condition for
$n=N$.  At $n=\frac{1}{2}N$, the appropriate boundary condition is
$\tau_{N/2}=0$; namely, the balanced state corresponds to the end of the
process. The result for $\tau_n$ for arbitrary $n$ is given in
\eqref{taun-final-app}, while the return time $RT$ itself is given in
\eqref{t1pn2} (see \ref{app:RT} for details).  Figure~\ref{RT} shows that
$RT$ scales as $\lambda^{-1}$ for $\lambda\to 0$.  This behavior is the
source of the long-lived temporal fluctuations in the composition of the
population, as illustrated in Fig.~\ref{fig:traj}.

\section{Discussion}
\label{sec:disc}

Much of the literature on populations of multiple cooperating species has
focused on the continuous limit.  Such a population is driven to an attractor
state in which there are equal concentrations of each species.  However, when
finite-population stochasticity is incorporated, the true attractors of the
dynamics of an isolated population are instead the fixation states, in which
only one species exists.  Because stochastic effects are relevant in real
systems, a discrete approach that incorporates this stochasticity is
necessary to describe the dynamics in a faithful way.

Within the discrete approach, we determined the probability for a finite
population to reach a given fixation state as a function of the initial
condition, as well as the time to reach fixation.  The behaviors of these two
quantities reflect the effective bias that drives the system to the
quasi-steady state of equal concentrations of the two species.  Namely, the
fixation probability is nearly independent of the initial condition and the
fixation time scales exponentially with population size $N$.  As a
consequence of this exponential dependence, fixation is not observable in a
laboratory or in a simulational time scale for any reasonable population
size.

It is worth mentioning that the statistical features of two-species
cooperation share similarities with the vacillating voter
model~\cite{lambiotte2007dynamics}, despite their different microscopic
update rules.  In the vacillating voter model, agents (voters) can be in one
of two voting states and their agreement or disagreement is influenced by the
state of yet another neighbor.  Here ``vacillation'' refers to the
possibility that a voter does not adopt the state of a randomly chosen
neighbor, as in the standard voter model, but may adopt the state of this
additional neighbor.  The properties of this decision process drives the
population toward a zero-magnetization state.  This state is equivalent to
the attractor in two-species cooperation, where both species are equally
represented.

When migration into the system can also occur, the population now ostensibly
reaches a steady state.  An unanticipated feature of this steady state for
small migration rate $\lambda$ is that this state is not genuinely steady,
because there are long-term stochastic fluctuations that drive the population
from one near-fixation state to the opposite near-fixation state, with the
population spending long time periods in these near-fixation states.  Such
macroscopic wanderings are reflected in the steady-state abundance
distribution, which is strongly peaked at these near-fixation states for
sufficiently small $\lambda$.  The time scale associated with these
fluctuations increases rapidly as the migration rate decreases.  Thus
observations of a cooperative system have to be sufficiently long to
incorporate many such wanderings so as to ensure that true average behavior
is actually probed.  We found that the return time---the time between
successive instants where the number of A's and B's are equal---allows us to
quantify these temporal fluctuations in a precise way.

This observation of large fluctuations also has important ramifications for
populations that consist of more than two cooperating species.  Depending the
immigration rate, the population size $N$, and the number of distinct species
$S$, the number of species that are actually present at any given time could
be much less than $S$.  Thus a \emph{typical} steady state could have a very
different character than the \emph{average} steady state that is predicted by
the time-independent density distribution.  Moreover, a multispecies
population will also exhibit large fluctuations in the actual composition of
the species that are present.  This intriguing issue has recently been
investigated in the context of multispecies Lotka-Volterra
models~\cite{bunin2017ecological,pearce2020stabilization}.  The backward
Kolmogorov equation offers the possibility of obtaining new insights into
these large fluctuations because of the relative simplicity of neutral models
of cooperating species.

The predictions presented in our study can also be tested in experimental
settings that are based on microfluidic chambers, where small populations of
cells can be maintained so that long-term monitoring can be performed.  These
small-scale devices offer a unique opportunity to explore the impact of
population size and validate the approximations presented in this work.  Both
well-mixed populations, as well as two- and there-dimensional spatial
populations, could be maintained in constant numbers over time.  Future
extensions of our model, such as including cell death or environmental noise,
would be helpful to design experimental protocols and explore the conditions
required to maintain stable cooperative cell
assemblies~\cite{ferry2011microfluidics,luke2016microfluidic}.

\subsection*{Acknowledgments}

The authors thank Deepak Bhat and Jacopo Grilli for fruitful discussions, and
Eric A. Blair for inspiring ideas. JP and RS were supported by the Bot\'in
Foundation and the Spanish Ministry of Economy and Competitiveness through
grant FIS2015-67616-P MINEICO/AEI/FEDER.  JP is also supported by "Mar\'ia de
Maezt\'u" fellowship MDM-2014-0370-17-2.  SRs research was supported in part
by NSF grant DMR-1910736.  JP and RS thank the hospitality of the Santa Fe
Institute, where this project began.

\appendix

\section{The Fixation Probability}
\label{app:En}

We want to solve Eq.~\eqref{En} for the fixation probability:
\begin{align*}
  E_n= a_n E_{n+1} + b_n E_{n-1}  +\big(1-a_n-b_n\big)E_n\,.
\end{align*}
This calculation is standard (see, e.g., \cite{karlin2014first}) and we
provide it here so that our presentation is self contained.  It is convenient
to rewrite the above equation as $a_n(E_{n+1}-E_n) = b_n(E_n-E_{n-1})$, and
then define $u_n\equiv E_n-E_{n-1}$ and $r_n\equiv b_n/a_n$ to recast it as
the following first-order recursion for the $u_n$:
\begin{align*}
  u_n=r_{n-1}\,u_{n-1}= r_{n-1}\,r_{n-2}\,r_{n-2}\cdots r_1\, u_1\,,
\end{align*}
with $u_1=E_1-E_0=E_1$.  We now define $R_n=\prod_{m=1}^n r_m$ so that the
equation for $u_n$ becomes
\begin{align}
  \label{un}
u_n=R_{n-1}\,u_1= R_{n-1}\,E_1\,.
\end{align}

Since the $u_n$ are successive differences of the $E_n$, we determine $E_n$
by summing the $u_n$.  Thus
\begin{align}
  \label{En-app}
  E_n=\sum_{m=1}^n u_m=\sum_{m=1}^n R_{m-1}\, E_1 = \sum_{m=0}^{n-1} R_m\, E_1\,,
\end{align}
where we need to define $R_0=1$ for consistency.  We now determine the
unknown $E_1$ by using the boundary condition $E_N=1$ in \eqref{En-app} to
give $E_N= \sum_{m=0}^{N-1} R_m\, E_1=1$.  Having found $E_1$, the general
solution for $E_n$ is
\begin{align}
\label{En-app-soln}
  E_n = \sum_{m=0}^{n-1} R_m \,\Big/ \sum_{m=0}^{N-1} R_m \,,
\end{align}

To simplify the above expression, we start with $a_n$ and $b_n$ defined in
Eq.~\eqref{ab}:
\begin{align*}
  a_n= \frac{n}{N}\left(1-\frac{n}{N}\right)^2\qquad
  b_n= \left( \frac{n}{N}\right)^2\left(1-\frac{n}{N}\right)\,,
\end{align*}    
so that $r_n=b_n/a_n= n/(N-n)$.  Then
\begin{align*}
  R_n&= r_1\,r_2\,\ldots r_n = \frac{n!(N-n-1)!}{(N-1)!}
       = \left[\binom{N-1}{n}\right]^{-1}\,.
\end{align*}
Substituting this representation for $R_n$ in \eqref{En-app-soln} gives
Eq.~\eqref{En-soln}.

For completeness, we also give the continuum solution for the fixation
probability.  We take the continuum limit of Eq.~\eqref{En} by letting
$(n\pm 1)/N\to x\pm dx$, with $dx=\frac{1}{N}$, and then expanding this
equation to second order in $dx$.  This gives
\begin{equation}
  \label{E-cont}
  E''+2N(1-2x)E'=0\,,
\end{equation}
where the prime denotes differentiation with respect to $x$.  This equation
is subject to the boundary conditions $E(0)=0$ and $E(1)=1$.  As in the
discrete formulation, the first condition corresponds to the impossibility of
reaching a population of all A's if the initial state contains no A's, while
the second condition corresponds to the initial state coinciding with the
desired final state.  The solution to \eqref{E-cont}, subject to the given
boundary conditions is
\begin{align}
  \label{Ex-soln}
  E(x) = \frac{{\displaystyle\int_0^x du \,e^{2N(u^2-u)}}}
  {{\displaystyle\int_0^1 du \,e^{2N(u^2-u)}}}
  =\frac{1}{2}\left[1+\frac{\text{erfi}\big(\sqrt{2N}(x-\frac{1}{2})\big)}
  {\text{erfi}(\sqrt{N/2})}\right]\,,
\end{align}
where erfi is the imaginary error function.  This expression from the
continuum approximation agrees well with the exact discrete result
\eqref{En-soln} (see Fig.~\ref{En-Ex}).  However, the continuum approach is
no longer accurate for the fixation time (see the next section).

\section{The Fixation Time}
\label{app:t}

We now solve the recursions \eqref{Tn} for
the fixation time:
\begin{align*}
  T_n&= a_nT_{n+1}+b_nT_{n-1}+\big(1-a_n-b_n\big)T_{n}+\delta t\,.
\end{align*}
Following the same steps that led to \eqref{un}, we obtain, for the
difference $v_n\equiv T_n-T_{n-1}$,
\begin{align}
\label{vn}  
  v_n= r_{n-1}\,v_{n-1}- \alpha_{n-1}\,,
\end{align}
where $\alpha_n\equiv \delta t/a_n$.  Notice that
$v_1= T_1-T_0=T_1 \equiv R_0\, T_1$.

We develop the recursion \eqref{vn} to give
\begin{align*}
  v_n &= r_{n-1}\,r_{n-2}\, v_{n-2} - \alpha_{n-1}-r_{n-1}\,\alpha_{n-2}\\
      &= r_{n-1}\,r_{n-2}\, r_{n-3}\, v_{n-3} - \alpha_{n-1}-r_{n-1}\,\alpha_{n-2}
        -r_{n-1}\,r_{n-2}\,\alpha_{n-3}\\
     &\hspace{3cm}\vdots
\end{align*}
continuing this development to the end leads to
\begin{align}
  \label{vn-end}
  v_n = R_{n-1} \,v_1 - Q_{n-1} = R_{n-1}\, T_1 - Q_{n-1}\,,
\end{align}
where
\begin{align*}
  Q_{n} \equiv  \alpha_{n}+r_{n}\,\alpha_{n-1}+r_{n}\,r_{n-1}\,\alpha_{n-2}+\cdots+
  r_{n}\, r_{n-1}\cdots r_2\, \alpha_1\,,
\end{align*}
and $Q_0=0$ by virtue of \eqref{vn-end}.

Finally, we sum the $v_n$ to obtain the fixation time:
\begin{align}
  \label{tn1}
  T_n =\sum_{m=1}^n v_m = \sum_{m=0}^{n-1}  R_{m}\; T_1 - \sum_{m=1}^{n-1}Q_m
\end{align}
To eliminate the unknown $T_1$, we use the boundary condition $T_N=0$ to give
\begin{align*}
  T_1 = \sum_{m=1}^{N-1} Q_m\Big/\sum_{m=0}^{N-1} R_m\,.
\end{align*}
Substituting this in \eqref{tn1} and noting that
$\sum_{m=0}^{n-1} R_{m}/ \sum_{m=0}^{N-1} R_{m}$ is just the fixation
probability $E_n$, we obtain the result quoted in Eq.~\eqref{tn-final}.

\section{WKB Approximation}
\label{app:WKB}

A comprehensive review of the application of the WKB method for large
deviations in stochastic populations can be found in \cite{assaf2017wkb}. In
this section we summarize the basic steps to reach Eq.~\eqref{S0_S1}. We make
the ansatz
$\tilde{P}(x)\sim \exp\left\{NS_0(x)+S_1(x)+O\left(N^{-1}\right)\right\}$,
and expand to linear order in $\delta x$ to give
\begin{align*}
  \tilde{P}(x\pm \delta x) \simeq \tilde{P}(x)\exp\left\{\pm
  S_0^{\prime}+ \delta x\left(\frac{S_0^{\prime\prime}}{2} \pm
    S_1^{\prime}\right)\right\}\simeq \ \tilde{P}(x)e^{\pm
  S_0^{\prime}}\left[1+\delta x\left(\frac{S_0^{\prime\prime}}{2} \pm
  S_1^{\prime}\right)\right]\,,
\end{align*}
with $\delta x = 1/N$.  We now define $\Lambda\equiv e^{S_0^{\prime}}$ and
substitute this into \eqref{quasisteady_process} to obtain
\begin{align*}
  \left(a-a^{\prime}\delta x\right) \Lambda^{-1} \left[ 1+\delta x
  \left(\frac{S_0^{\prime\prime}}{2} -
  S_1^{\prime}\right)\right]+\left(b+b^{\prime}\delta
  x\right)\Lambda\left[1+\delta x \left(\frac{S_0^{\prime\prime}}{2} +
  S_1^{\prime}\right)\right]=a+b\,,
\end{align*}
which can be separated into terms of $O(1)$ and terms of $O(\delta x)$. For
the former, we have
\begin{eqnarray}
  a\Lambda^{-1}+b\Lambda=a+b\,,
\end{eqnarray}
which has the two solutions $\Lambda_0=1$ and $\Lambda=\frac{a}{b}$.  The
$\Lambda_0$ solution corresponds to $S_0^{\prime}=0$, and the resulting
constant can be absorbed by the normalization condition on
$\tilde{P}(x)$.   The second solution is
\begin{eqnarray}
S_0^{\prime}=\log\left(\frac{a}{b}\right) \,,
\end{eqnarray}
which, after integration, results in the first expression in
\eqref{S0_S1}.   For the $O(\delta x)$ terms we must solve
\begin{eqnarray}
-a^{\prime}\Lambda^{-1}+a\Lambda^{-1}\left(\frac{S_0^{\prime\prime}}{2} - S_1^{\prime}\right) +b^{\prime}\Lambda+b\Lambda\left(\frac{S_0^{\prime\prime}}{2} + S_1^{\prime}\right)=0 \,,
\end{eqnarray}
which, after substitution of $\Lambda=\frac{a}{b}$, yields
\begin{eqnarray}
b\left(\frac{S_0^{\prime\prime}}{2} - S_1^{\prime}\right)+a\left(\frac{S_0^{\prime\prime}}{2} + S_1^{\prime}\right)=b\frac{a^{\prime}}{a}-a\frac{b^{\prime}}{b} \;.
\end{eqnarray}
On the other hand,
$S_0^{\prime\prime}=\frac{a^{\prime}}{a}-\frac{b^{\prime}}{b}$, which reduces
the previous equation to
\begin{eqnarray}
  (a-b)S_1^{\prime}=\frac{1}{2}\left[b\;\frac{a^{\prime}}{a}-a\;\frac{b^{\prime}}{b}
  -(a^{\prime}-b^{\prime})\right]
=-\frac{1}{2}\left(\frac{a^{\prime}}{a}+\frac{b^{\prime}}{b}\right)(a-b) \ .
\end{eqnarray}
The first term in the brackets on the right-hand side corresponds to the
derivative of $\log(ab)$.  Hence, after integration, we obtain the second
expression in \eqref{S0_S1}.

\section{The Return Time}
\label{app:RT}

We want to solve the recursion for the return time $\tau_n$, defined as the
time for the population to first reach the state with equal numbers of A's
and B's when the initial state contains $n>\frac{1}{2}N$ A's.  The state with
$n=N$ plays the role of an effective reflecting boundary condition.  The
system of equations that we wish to solve is \eqref{taup}:
\begin{subequations}
\begin{align}
  \label{taup-app}
\tau_n= a_n'\tau_{n+1} +b_n'\tau_{n-1} +(1-a_n'-b_n')\tau_n+\delta t\,,
\end{align}
for $1+N/2\leq n<N$, while for $n=N$ the appropriate equation is
\begin{align}
  \tau_N= b_B'\tau_{N-1} +(1-a_N'-b_N')\tau_N+\delta t\,.
\end{align}
\end{subequations}
Using the fact that $a_N'=0$, this last equation can be rewritten as
\begin{align}
  \label{dtN-ref}
  \tau_N-\tau_{N-1} \equiv v_N =\frac{\delta t}{b_N'}\,.
\end{align}

The remaining $\frac{N}{2}-1$ equations \eqref{taup-app} are of the same type
as \eqref{Tn} for the fixation time.  Thus the solution for $\tau_n$ has
the same form as \eqref{tn1}:
\begin{align}
  \label{taun1-app}
  \tau_n = \sum_{m=N/2}^{n-1}  R_{m}'\; \tau_{1+N/2} \;-\!\! \sum_{m=1+N/2}^{n-1}Q_m'\,,
\end{align}
where 
\begin{align*}
  R_n'=\prod_{m=1+N/2}^n \frac{b_n'}{a_n'} \equiv \prod_{m=1+N/2}^n r_n'
\end{align*}
and
\begin{align*}
  Q_{n}'= \alpha_{n}'+r_{n}'\,\alpha_{n-1}'+r_{n}'\,r_{n-1}'\,\alpha_{n-2}'+\cdots+
  r_{n}'\, r_{n-1}'\cdots r_{2+N/2}'\, \alpha_{1+N/2}'\,,
\end{align*}
with $\alpha_n'=\delta t/a_n'$ and and $Q_{N/2}'=0$.

To eliminate the unknown $\tau_{1+N/2}$, we now write \eqref{taun1-app} for
the special cases of $n=N$ and $n=N-1$:
\begin{align*}
  \tau_N = \sum_{m=N/2}^{N-1}  R_{m}'\; \tau_{1+N/2}\; - \!\!\sum_{m=1+N/2}^{N-1}Q_m'\qquad\qquad
   \tau_{N-1} = \sum_{m=N/2}^{N-2}  R_{m}'\; \tau_{1+N/2}\; - \!\!\sum_{m=1+N/2}^{N-2}Q_m'\,.
\end{align*}
Their difference is
\begin{align*}
  \tau_N-\tau_{N-1}=R_{N-1}'\, \tau_{1+N/2}-Q_{N-1}' = \frac{\delta t}{b_N'}\,,
\end{align*}
so that $\tau_{1+N/2}$ is given by
\begin{align}
\label{t1pn2}
  \tau_{1+N/2} = \frac{Q_{N-1}'+\delta t/b_N'}{R_{N-1}'}\,.
\end{align}
Substituting this expression for $\tau_{1+N/2}$ in \eqref{taun1-app} gives
the average time to reach the balanced state of equal numbers of A's and B's
when starting from a population that contains $n$ A's with a reflecting
boundary condition at $n=N$:
\begin{align}
  \label{taun-final-app}
  \tau_n = \sum_{m=N/2}^{n-1} R_{m}'\;  \frac{Q_{N-1}'+\delta t/b_N'}{R_{N-1}'}
  - \sum_{m=1+N/2}^{n-1}Q_m'\,.
\end{align}
What we want, however, is, the return time, defined as the average time to
start at the balanced state and first return to this state.  This is
$RT=\frac{1}{N}+\tau_{1+N/2}$.

\newpage


\providecommand{\newblock}{}

\end{document}